# Observation of atomically displacive transformation out of the boundary-reconstructive phase competition

Qingqi Zeng[1], Zhiwei Du[2], Xiaolei Han[2], Binbin Wang[1], Guangheng Wu[1], Enke Liu[1*]

1. Beijing National Laboratory for Condensed Matter Physics, Institute of Physics, Chinese Academy of Sciences, Beijing 100190, China.
2. Guobiao (Beijing) Testing & Certification Co., Ltd., Beijing 100088, China.

E-mail: ekliu@iphy.ac.cn

## Abstract

During the phase transitions, diverse states evolve with multiplex phenomena arising from the critical competition. In this study, a displacive martensitic transformation with a lattice shear distortion was unexpectedly observed at the reconstructive phase boundary that usually connects multiple phases without crystallographic relation, in a Ni–Co–Mn–V all-$d$-metal alloy system. Experiments and theoretical calculations suggest that the parent phase becomes increasingly unstable when approaching the phase boundary. The lattice-distorted transformation with moderate first-order nature survives due to the critical phase competition from the structural frustration, in which the comparable energy and the diminished formation preference of different phases emerge. In this critical state, the phase selection including the martensitic phase transformation can be tuned by external fields such as rapid cooling, annealing and magnetic field. Our research reveals a novel manner to destabilize the parent phase, through which one could attain new functional materials based on the phase transitions.

# 1. Introduction

    Various forms of phase transitions have been observed in condensed matter, including liquid–liquid, solid–liquid, atomically diffusive solid-state, atomically displacive solid-state, and higher-order electronic phase transitions[1, 2]. All these transitions prefer to eliminate the instability of the initial system and reach a more stable resultant one. Martensitic transformation (MT)[3, 4] is an atomically displacive first-order structural phase transformation with latent heat release. It occurs when external conditions such as temperature, stress, or magnetic field alter. A key feature of MTs is the existence of a crystallographic relation between the two phases and the initial lattice shears to a resultant distorted with atom motion within a unit cell, which lowers the energy barrier of the structural phase transition[5, 6]. In contrast, reconstructive structural transformation is characterized by a long-range-atomic diffusion. These two phase transitions can be described by Landau theory, under which one can use an order parameter to describe their local symmetry breaking during the phase transitions.

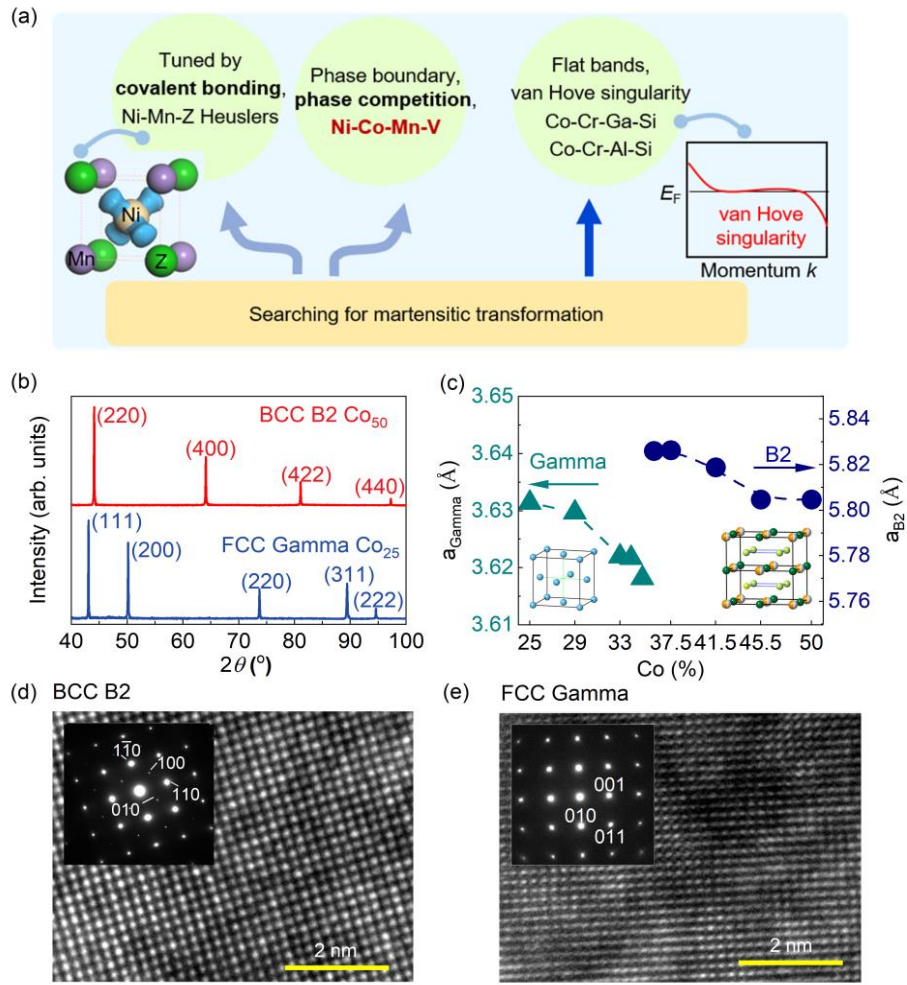

Fig. 1. (a) Ways of searching for martensitic transformations (MT). Insets are illustrations of covalent bonding (left) and van Hove singularity (right), respectively. (b) XRD patterns of $Ni_{25}Co_{25}Mn_{25}V_{25}$ ($Co_{25}Mn_{25}$) and $Co_{50}Mn_{25}V_{25}$ ($Co_{50}Mn_{25}$) ribbon samples measured at room temperature. BCC and FCC indicate body-center-cubic and face-center-cubic lattices, respectively. (c) Lattice parameters versus Co content. Insets show conventional lattices. (d–e) High-resolution transmission electron microscope (HRTEM) and corresponding selected area electron diffraction (SAED) patterns viewed along [001] and [100] zone axes for $Co_{50}Mn_{25}$ (d) and $Co_{25}Mn_{25}$ (e) samples, respectively.

MTs were intensively reported in Heusler[7-9], MM'X[10], and all-$d$-metal Heusler[9, 11] alloys. Abrupt changes in the structure, shape, magnetism, and release of latent heat during an MT are applicable to multiple functional and caloric materials[4, 12-20]. The basics of the aforementioned applications are various MT materials. To search for MT, there are the following ways. In experiments, many materials with MT are realized by adjusting the covalent bonding (illustrated in Fig. 1(a)). As a representative case, the binary NiMn alloy exhibits a B2-to-L1$_0$ MT at high temperatures[21]. By introducing main-group elements or low-valence transition metals into this system[11, 22], the MT can be tailored to lower temperatures. This is because the main-group elements and low-valence transition metals form $p$–$d$ or $d$–$d$ covalent bonds (left of Fig. 1(a)) and stabilize the parent phase to lower temperature. In theoretical research works, the possibility of MT in a material is linked to the high density of states (DOS) value, i.e., the van Hove singularity at Fermi level (right of Fig. 1(a))[23-28]. Although MT materials are intensively reported, constructing new systems and finding diverse physical responses during the transformation are still of great interest. However, only one way in experiments to find MT will limit the construction of new MT materials.

In this study, we report a new method to search for MTs. It is found that the MT in Co-doped $Ni_{50}Mn_{25}V_{25}$ all-$d$-metal Heusler alloys originates from the phase competition at the reconstructive phase boundary in this system. The compositions with MT properties are found in a narrow composition region separating the B2 and gamma phases. In the following, we will exhibit our experimental and calculated results to elucidate that the displacive transformation originates from the competition of forming different structures. By utilizing the competition, one can destabilize a B2 structure, through which potential new ferroic and caloric materials can be constructed.

## 2. Materials and methods

### 2.1 Sample preparation

Polycrystalline precursor ingots were prepared by arc-melting high-purity metals in an argon atmosphere. The ingots were melted four times and turned over in between to guarantee homogeneity. The as-prepared ingots were subsequently melt-spun in an argon atmosphere onto a copper wheel traveling at a wheel linear speed of 20 m s$^{-1}$. The obtained melt-spun ribbons are approximately 2 mm wide and 50 μm thick.

### 2.2 Structure and morphology

Room temperature X-ray diffraction was performed on the melt-spun ribbons by using a Rigaku XRD D/max 2,400 diffractometer with Cu-$K\alpha$ radiation. The morphology of martensite variants was observed by the scanning electron microscope (JEOL JSM-7001F). Transmission electron microscope (FEI Tecnai G2 F20) was employed to obtain the selected area electron diffraction and high-resolution transmission electron microscope patterns.

### 2.3 Measurements for physical properties

A Quantum Design physical property measurement system (PPMS) was used for magnetic and electric transport measurements with a cooling and heating rate of 5 K min$^{-1}$. The resistance is measured by using the four-point method with a constant electric current of 5 mA. Thermal analysis was performed using differential scanning calorimetry (DSC, NETZSCH) with a cooling and heating rate of 10 K min$^{-1}$. The martensitic and magnetic transformation temperatures were determined by DSC and magnetic measurements.

### 2.4 Theoretical calculations

The electron-structure and magnetic moments were calculated using the Korringa-Kohn-Rostoker method combined with the coherent potential approximation (KKR-CPA method) [29]. Formation energy calculation was performed within the density functional theory (DFT) using the Cambridge serial total energy package (CASTEP) [30]. The exchange-correlation energy in the generalized gradient approximation (GGA) of Perdew was adopted. Further, the effect of doping with Ni was taken into account by reducing the symmetry of the structure and substituting Co atoms.

## 3. Results

Upon substitution of Co for Ni in $Ni_{50-x}Co_xMn_{25}V_{25}$, gamma (face-centered-cubic, FCC) and B2 (body-centered-cubic, BCC) type structures are observed with low-Co content (<35 at.%) and high-Co content (>36 at.%), respectively. The representative X-ray diffraction (XRD) patterns of B2 and gamma phases are shown in Fig. 1(b). For both phases, the lattice parameter decreases with increasing Co content, as shown in Fig. 1(c) (see Supplementary Materials Note 1 for further information on the basic parameters of this system). Conventional lattices of both phases are shown in the insets of Fig. 1(c). Figs. 1(d–e) show the high-resolution transmission electron microscope (HRTEM) patterns, and the insets show the selected area electron diffraction (SAED) patterns viewed along [001] and [100] zone axes. The diffraction spots can be indexed as the B2 and gamma phases, respectively, which are consistent with the XRD results.

Figure 2(a) shows the thermo-magnetization ($M$–$T$) curves in a magnetic field of 0.1 kOe of $Ni_{12.5}Co_{37.5}Mn_{25}V_{25}$ ($Co_{37.5}Mn_{25}$), $Ni_{17}Co_{33}Mn_{25}V_{25}$ ($Co_{33}Mn_{25}$) and $Ni_{14}Co_{36}Mn_{25}V_{25}$ ($Co_{36}Mn_{25}$) samples. Both $Co_{37.5}Mn_{25}$ and $Co_{36}Mn_{25}$ are B2 structures, while $Co_{33}Mn_{25}$ is gamma. The magnetization of $Co_{37.5}Mn_{25}$ increases up to 100 emu g$^{-1}$ with decreasing temperature below the Curie temperature ($T_C$ ~ 350 K). $Co_{33}Mn_{25}$ exhibits a very low magnetization over the entire range of measuring temperature. For $Co_{36}Mn_{25}$ (the one near the reconstructive phase boundary), apart from the increase in magnetization just below $T_C$, an unexpected sudden drop occurs at 254 K upon further cooling, indicating a temperature-driven phase transformation. The hysteresis in the $M$–$T$ curve upon heating indicates the first-order nature of this transformation, corresponding to the potential MT at such low temperature in a solid. The parent phase has high magnetization while the martensite phase has low magnetization, resulting in a large difference ($\Delta M$) of 50 emu g$^{-1}$ in 0.1 kOe during the transformation.

Fig. 2(b) shows the thermal measurements (lines) on $Co_{36}Mn_{25}$ by differential scanning calorimetry (DSC), in which the cooling and heating cycles are indicated by arrows. DSC curves exhibit exothermic and endothermic peaks. The transformation starting ($M_s$) and finishing ($M_f$) temperatures are 254 and 230 K, respectively. Moreover, the reverse transformation starting ($A_s$) and finishing ($A_f$) temperatures are 317 and 363 K, respectively. A thermal hysteresis ($A_f - M_s$) of 109 K is observed, which indicates the first-order nature of the MT. As an MT is always accompanied by a resistivity change, the temperature dependence of the resistivity was measured in zero field (Fig. 2(b), dots). Two abnormal changes (indicated by arrows) in the resistivity correspond to the forward and reverse MTs, which are consistent with the DSC measurements. One can see that martensite has a lower resistivity. However, because of the lattice deformation during the MT, the resistance of almost all ferromagnetic shape memory alloys[31, 32] increases after the MT except $Fe_2MnGa$[33] and $Mn_2NiGa$[34]. This rarely observed behavior in this study may be attributed to the abrupt change in the electronic structure during the MT[33].

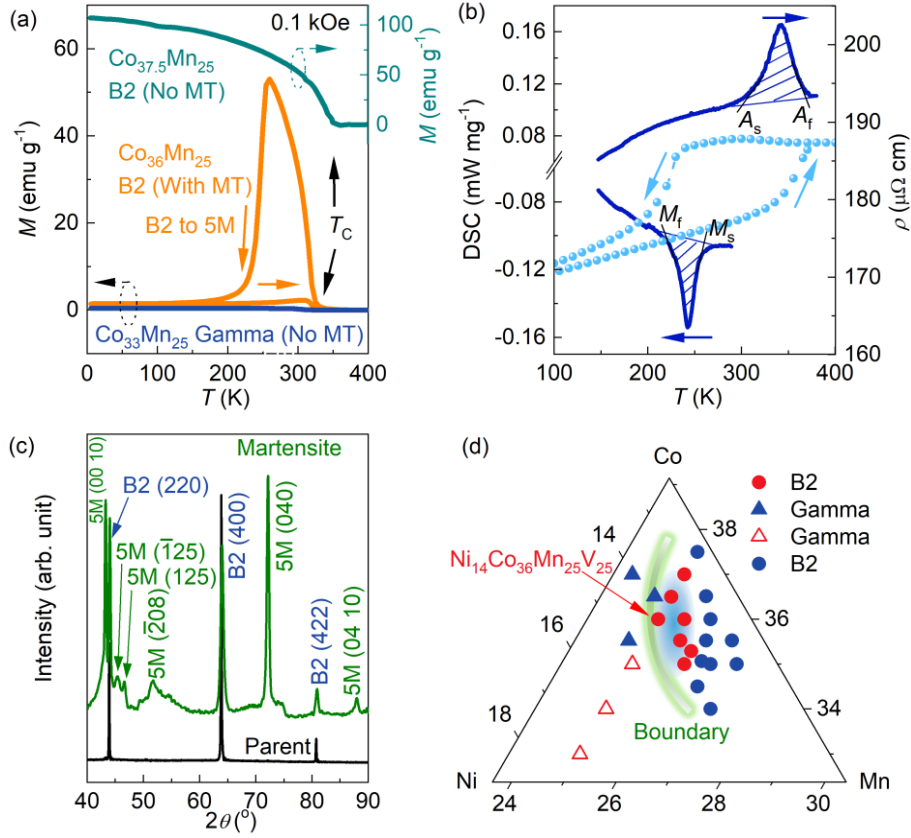

Fig. 2. (a) *M–T* curves of $Ni_{17}Co_{33}Mn_{25}V_{25}$ ($Co_{33}Mn_{25}$), $Ni_{14}Co_{36}Mn_{25}V_{25}$ ($Co_{36}Mn_{25}$), and $Ni_{12.5}Co_{37.5}Mn_{25}V_{25}$ ($Co_{37.5}Mn_{25}$) samples measured in a magnetic field of 0.1 kOe. (b) Temperature dependences of differential scanning calorimetry (DSC, left) and zero-field resistivity (right) of $Co_{36}Mn_{25}$ sample. Cooling and heating cycles are indicated by arrows. Transformation starting ($M_s$) and finishing ($M_f$) temperatures; and reverse transformation starting ($A_s$) and finishing ($A_f$) temperatures are denoted. (c) Room-temperature XRD patterns of parent (lower) and martensite (upper) phases of $Co_{36}Mn_{25}$ sample. Martensite states in (c) were obtained by soaking the sample in liquid nitrogen for 20 min. (d) Pseudo-ternary phase diagram (relative content of Ni, Co and Mn is normalized as $Ni_xCo_yMn_z$, where x + y + z = 75) of Ni–Co–Mn–V alloys. Triangle and dot symbols represent gamma and B2 phases, respectively. Red dots represent compositions with MT; red open triangle symbols indicate compositions with MT after heat treatments. Green line sketches phase boundary.

The MT in $Co_{36}Mn_{25}$ was further revealed by the room-temperature XRD patterns, as shown in Fig. 2(c). The sample retains the martensite state at room temperature after being warmed from low temperatures because the reverse transformation temperature is above room temperature. Hence, the martensite state was obtained by soaking the sample in liquid nitrogen for 20 min. For the parent phase, the characteristic diffraction peaks can be indexed to a B2-type structure. However, in the martensite phase, the diffraction peaks suggest a modulated structure. Similar to previous research on all-*d*-metal Heusler alloys[9, 11], the XRD pattern of the martensite phase can be indexed to a 5-layer modulated (5M) structure (with residual parent phase).

We collect the studied compositions to form a pseudo-ternary phase diagram, as shown in Fig. 2(d). The related parameters of MT can be found in Supplementary Materials Note 2. It is important to note that the diagram is not a cross-section of the $V_{25}$ composition. We normalized the relative contents of Ni,

Co and Mn to $Ni_xCo_yMn_z$, where $x + y + z = 75$. The samples with MT properties (red dots) are found in a narrow composition region, which is located in the boundary separating the B2 and gamma phases.

We subsequently employed a scanning electron microscope and a transmission electron microscope to probe the morphology and structure of $Co_{36}Mn_{25}$. The results show a sharp contrast to the samples without MT (Figs. 1(d–e)). Figure 3(a) shows the spherical-grain morphology (left) and SAED patterns (right) of the parent phase. The SAED results show that the parent phase has the cubic B2-type structure, which is consistent with the XRD results. Fig. 3(b) shows the plate-like martensite shear variants with a plate thickness of 0.5 to 1 μm (left) and the corresponding SAED patterns (right). Even though the diffraction spots are elongated in the martensite phase, we can identify four satellites between the two main reflections. This indicates that the 5M monoclinic structure with the space group of P2/m is the main part of the martensite phase. The HRTEM pattern of the martensite phase in Fig. 3(c) reveals that the shear modulation in the martensite phase is complex and multi-substructures. This irregularity can account for the elongated satellites in the SAED pattern. The orange lines in the enlarged inset indicate a 5-fold modulated periodic structure within a small region. To show the lattice relation, we depict the structure of the 5M martensite and the distorted B2-type parent structure (red lines) in Fig. 3(d).

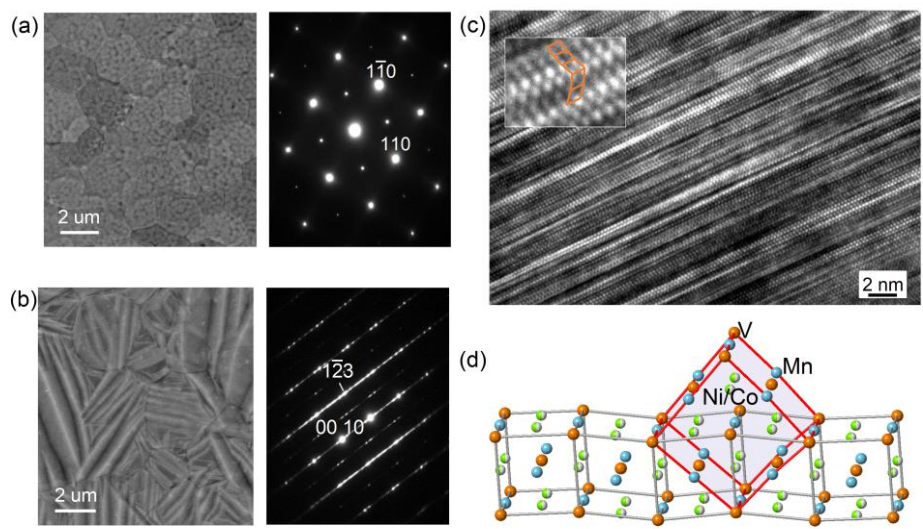

Fig. 3. (a) and (b) Backscattered electron (left) and SAED (right) patterns of parent phase and martensite variants of $Co_{36}Mn_{25}$, respectively. SAED patterns are viewed along [001] and [210] zone axes, respectively. (c) HRTEM of martensite phase of $Co_{36}Mn_{25}$ sample. Inset shows an enlarged part. Orange lines provide visual guidance. (d) Structure of 5-fold modulated (5M) martensite with a distorted B2 parent lattice shown by red lines.

In Fig. 1, we have illustrated the notable structural differences between the two phases separated by this boundary. Additionally, their magnetic properties exhibit a sharp contrast. The magnetic moments ($M$) measured at 5 K versus Co content of the $Ni_{50-x}Co_xMn_{25}V_{25}$ ribbon samples are shown in Fig. 4(a) (see magnetization curves in Supplementary Materials Note 1). Samples with gamma and B2 phases are weak magnetic and ferromagnetic, respectively. The measured moments of the B2 phase are in agreement with the calculated results based on ferromagnetic order. However, the $Co_{36}Mn_{25}$ composition is an exception as MT occurs near 250 K and is in the martensite state at 5 K.

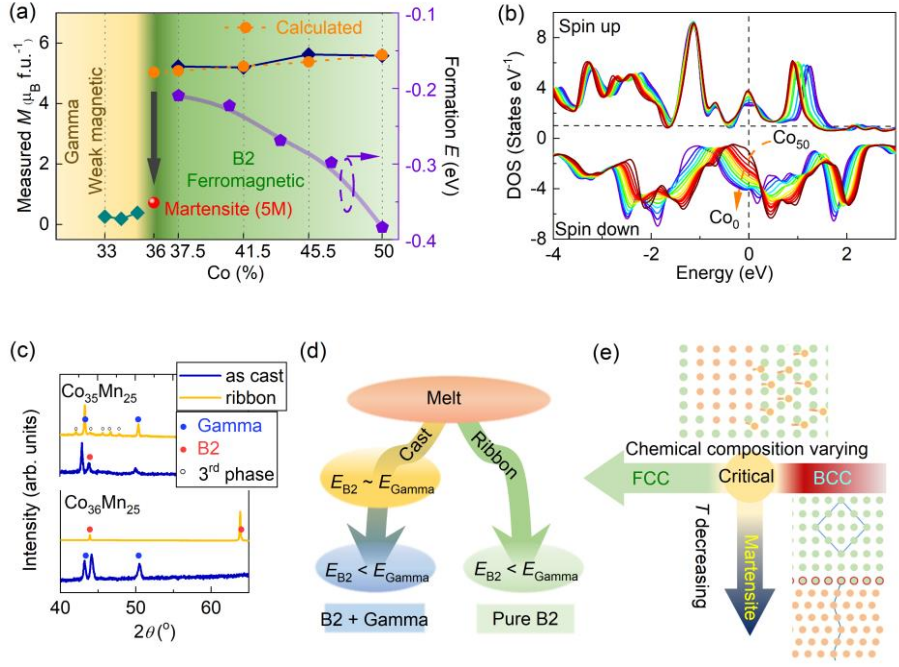

Fig. 4. (a) Magnetic moments ($M$) of $Ni_{50-x}Co_xMn_{25}V_{25}$ ribbon samples, which are characterized by saturation magnetization measured at 5 K. Orange dots indicate the calculated results by KKR-CPA method (left axis). Co content dependence of the formation energy calculated by CASTEP (right axis). (b) Calculated spin-projected DOS for the B2-type $Ni_{50-x}Co_xMn_{25}V_{25}$ (denoted as $Co_x$) systems. (c) XRD data of as-cast and ribbon samples for different compositions. (c) XRD data of as-cast and ribbon samples for different compositions. (d) Sketch of sample preparation. (e) Schematic illustration of phase competition around phase boundary. Upper inset sketches atomic diffusion. Lower right inset shows crystallographic relation between parent and martensite phases. There exists a habit plane between two phases.

## 4. Discussion

Next, we provide a possible interpretation of why the MT occurs at the phase boundary. It is widely accepted that the high value of DOS at $E_F$ in the cubic parent phase indicates a probability of a lattice distortion[25-28]. Fig. 4(b) shows the calculated DOS near the Fermi energy ($E_F$) for $Ni_{50-x}Co_xMn_{25}V_{25}$ compositions with B2-type structure. With increasing Co content, the DOS peak near the $E_F$ in the spin-up channel changes little. In contrast, the $E_F$ shifts to a higher DOS region in the spin-down channel. As a consequence, the total DOS at the Fermi energy ($E_F$) increases with decreasing Co content, suggesting that the B2 structure becomes increasingly unstable as it approaches the phase boundary. [35].

The right axis in Fig. 4(a) shows the Co-dependence of the formation energy calculated using the Cambridge serial total energy package (CASTEP). The details of the calculations can be found in Materials and Methods and Supplementary Materials Note 3. The negative formation energy values of the B2 structure in Fig. 4(a) indicate that these compounds can be stably alloyed. However, as one can see, the absolute value of formation energy decreases with decreasing Co content, indicating that the system gradually becomes unstable near the reconstructive phase boundary. The B2 structure can still be formed when the Co content reaches 36%. However, the composition exhibits temperature-induced MT from B2 to 5M structures, indicating an instability that is consistent with the calculation results.

Additionally, it is challenging to form a B2 structure below 36 at.% Co content.

It should be kept in mind that the studied ribbon samples were prepared using a melt-spinning method with a high cooling rate. Without fast-cooling, the as-cast bulk samples prepared by the arc melting with a normal cooling rate contain both B2 and gamma phases near $Co_{36}Mn_{25}$ composition (Fig. 4(c)), which implies that the dual-phase region connects the pure B2 and gamma regions. The entanglement of the different phases can be suppressed by melt spinning. Then the samples with Co contents less and more than 36 at. % form gamma (with traced other phase) and B2 structures, respectively, leaving a sharp phase boundary with a dual-phase in a region of approximately 1% (Fig. 4(c)).

The instability of the B2 structure at phase boundary is unambitiously shown based on the high DOS and formation energy, and the entanglement of different phases. Next, we show an interpretation of the realization of MT. The stability of a phase is conventionally evaluated by thermodynamic energy. Low (high) Co content samples form the gamma (B2) phase as the B2 (gamma) phase possesses a higher energy level. Nevertheless, the sample preparation is not a thermodynamic equilibrium process. Owing to the structural frustration at the critical phase boundary, the multiple phases are comparable to each other in energy. With a diminished difference in energy between the two phases at the phase boundary, multiple phases can nucleate during a non-equilibrium solidification. In addition, these phases can hardly be eliminated during cooling to room temperature due to a lack of thermal-activated energy (sketched on the left side of Fig. 4(d)). Hence, phases that are not favorable in energy can also be retained to room temperature. In our studied system, over a wide composition range, the two phases coexist in the as-cast samples. During melt spinning, the liquid metal is rapidly cooled to room temperature. Owing to a relatively high driving force of crystallization in the fast-cooling process, the phase with thermodynamic preference survives (the supporting of B2 by rapid cooling is sketched on the right side of Fig. 4(d)). Specifically, melt spinning supports a B2 structure that is not absolutely favorable in energy. Therefore, its structure is sensitive to external parameters and the lattice-distorted MT with a moderate first-order nature thus occurs.

There are additional indications for the diminished preference of different structures at the phase boundary. The spinning-supported gamma structure contains a third phase, which is not the B2 or gamma structure (denoted by open dots in Fig. 4(c)). The B2 parent phase is supported by annealing at 673 K, which may be caused by a higher order of the atom occupation[36, 37]. By utilizing the heat treatment process, it is possible to find additional compositions that possess MT properties. Besides, a high magnetic field can also affect the relative phase stability and change the MT temperature due to strong magnetization of the B2 phase (see Supplementary Materials Note 4 for additional information). These behaviors further indicate that the phase competition and the survived phase transformation can be sensitively tuned by external conditions such as rapid cooling, annealing and magnetic field.

A schematic illustration is shown in Fig. 4(e) to exhibit the displacive transformation originating from the reconstructive phase boundary. Away from the phase boundary, stable BCC and FCC phases have no MT, between which an unstable B2 region exists. When the temperature is decreased, the unstable B2 phase undergoes an MT. The lattice distortion depicted in the lower right corner of Fig. 4(e) shows a sharp contrast with the atom diffusion transition (upper inset of Fig. 4(e)). Upon analyzing the previously reported alloys, we also noticed a material with a similar MT. The compositions of Fe-Mn-Ga Heusler alloys with MTs are also located at the phase boundary of B2 and gamma structures[33]. This example shows that similar phase competition may exist in other materials. Hence, utilizing the phase competition, like the case in our study, could be a general way to realize MTs. It is sufficient to suggest that other similar systems may exhibit the similar phase region around the phase boundary for MTs. A challenge, however, would be how to find such narrow composition ranges, which proposes a new topic for future investigations.

## 5. Conclusion

In this study, by employing a fast-cooling protocol, the dual-phase region in an all-d-metal Heusler alloy Ni–Co–Mn–V is reduced to a single-phase region. Just at the phase boundary separating the FCC and BCC phases, a displacive martensitic transformation surprisingly occurs out of the phase transition between FCC and BCC phases. With the increasing instability of the BBC parent phase, a diminished preference for forming different structures in the phase boundary is observed, which means that phase transitions can be induced by external parameters. The martensitic phase transformation is due to the phase competition due to the structural frustration at the critical phase boundary. The results indicate the phase competition and the survived phase transformation can be sensitively tuned in the system with such a phase boundary. This study demonstrates that the phase competition around the phase boundary not only reveals a new mechanism for martensitic phase transformation, but also offers an alternative method for discovering new phase-transition materials for potential applications such as shape memory, solid-state refrigeration, and energy conversion.

**Acknowledgments**

This work was supported by National Key R&D Program of China (No. 2019YFA0704900), Fundamental Science Centre of the National Natural Science Foundation of China (No. 52088101), National Key R&D Program of China (No. 2022YFA1403400), Beijing Natural Science Foundation (No. Z190009), National Natural Science Foundation of China (Nos. 11974394 and 51271038), the Strategic Priority Research Program (B) of the Chinese Academy of Sciences (CAS) (XDB33000000), the Synergetic Extreme Condition User Facility (SECUF), and the Scientific Instrument Developing Project of CAS (No. ZDKYYQ20210003).